\documentclass[aps,showpacs]{revtex4}
\usepackage[dvips]{graphicx}
\usepackage{amsmath}

\begin{document}
\title{Gauge procedure with gauge fields of various ranks}
\author{Douglas Singleton}
\email{dougs@csufresno.edu}
\affiliation{Centre of Gravitation and Fundamental Metrology, VNIIMS, 3-1 M.
Ulyanovoy St., Moscow 119313, Russia; \\
Physics Dept., CSU Fresno, 2345 East San Ramon Ave.
M/S 37 Fresno, CA 93740-8031}
\author{Akira Kato}
\email{ak086@csufresno.edu}
\affiliation{Physics Dept., CSU Fresno, 2345 East San Ramon Ave.
M/S 37 Fresno, CA 93740-8031}
\author{A. Yoshida}
\email{ayoshida@stab.org}
\affiliation{Science Dept., Saint Anne's Belfield School, Charlottesville, VA 22903}

\date{\today}

\begin{abstract}
The standard procedure for making a global phase symmetry local
involves the introduction of a rank 1, vector field in the definition of the
covariant derivative. Here it is shown that it is possible to gauge a
phase symmetry using fields of various ranks. In contrast to other formulations
of higher rank gauge fields we begin with the coupling of the gauge field to
some matter field, and then derive the gauge invariant, field strength tensor.
Some of these gauge theories are
similar  to general relativity in that their covariant derivatives
involve derivatives of the rank $n$ gauge field rather than
just the gauge field. For general relativity the covariant
derivative involves  the Christoffel symbols which are written
in terms of derivatives of the metric tensor. Many (but not all) of  the Lagrangians
that we find for these higher rank gauge theories lead to nonrenormalizable
quantum theories which is also similar to general relativity.
\end{abstract}


\maketitle

\section{Standard gauge procedure with a rank 1 field}

The concept of symmetries, and the process of turning global
symmetries into local ones ({\it i.e.} gauging the symmetry) are
important features of modern field theories. An
example is Maxwell's theory, which can
be derived from the gauge principle applied to an Abelian U(1) symmetry of
some matter field. In this section we summarize the standard gauge procedure.
For our matter field we will use a
complex scalar matter field, $\varphi$, throughout the paper.
The same procedure applies starting with other types of
matter field ({\it e.g.} a spinor field).
The Lagrange density for the matter field, $\varphi$, is
\begin{equation}
\label{1-1}
{\cal L} _{scalar} = (\partial _{\mu} \varphi )^{\ast}
(\partial ^{\mu} \varphi ) +...
\end{equation}
The ellipses leave off mass, $m^2 \varphi ^{\ast} \varphi$,
and self-interaction terms, $\lambda (\varphi ^{\ast} \varphi)^2$,
that don't involve derivatives of $\varphi$. This Lagrange density
satisfies the global phase symmetry
\begin{equation}
\label{1-2}
\varphi (x) \rightarrow e^{-ig \Lambda} \varphi (x) ~,
\; \; \; \;
\varphi ^{\ast} (x) \rightarrow e^{ig \Lambda } \varphi ^{\ast} (x) ~,
\end{equation}
where $g$ is the coupling and $\Lambda$ is a constant. This phase symmetry can be
made local ($g \Lambda \rightarrow g \Lambda (x)$) by replacing
the ordinary derivative with the covariant derivative
\begin{equation}
\label{1-3}
\partial _{\mu} \rightarrow D ^{[1]}_{\mu} \equiv
\partial _{\mu} -ig \sigma _{\mu \nu} A^{\nu} .
\end{equation}
Throughout the paper the bracketed superscript indicates the rank
of the gauge field. A rank-2 operator $\sigma_{\mu \nu}$ has been
introduced into this standard gauge procedure since this will
allow the generalization to the gauging procedure for different
rank gauge fields. The four-vector gauge field,
$A_{\mu}$, is required to transform as
\begin{equation}
\label{1-4}
A_{\mu} \rightarrow A_{\mu} - \Gamma_{\mu}.
\end{equation}
In order for the local symmetry version of \eqref{1-2} to be valid
$\Gamma_{\mu}$, $\sigma _{\mu \nu}$ and $\Lambda (x)$ must satisfy
\begin{equation}
\label{1-4a}
\sigma _{\mu \nu} \Gamma^{\nu} - \partial _{\mu} \Lambda =0.
\end{equation}
The standard case corresponds to choosing
\begin{equation}
\label{1-4b}
\sigma_{\mu \nu} = \eta _{\mu \nu} ~, ~~~~~~~~~~ \Gamma_{\mu} =
\partial _{\mu} \Lambda.
\end{equation}
 Another possible choice, which leads to a different theory, is
\begin{equation}
\label{1-4c}
\sigma_{\mu \nu} = \partial _{\mu} \partial _{\nu} ~, ~~~~~~~~~~
\Lambda = \partial _{\mu} \Gamma^{\mu} + f(x) ~,
\end{equation}
where $f(x)$ is a divergenceless function, $\partial _{\mu} f(x)
=0$. An important difference between the two options is that in
\eqref{1-4b} $\Gamma_{\mu}$ is defined in terms of the local phase
factor, $\Lambda (x)$, while for \eqref{1-4c} $\Lambda (x)$ is
defined in terms of $\Gamma_{\mu}$. For \eqref{1-4b} $\Lambda$ appears
more fundamental, while for \eqref{1-4c} $\Gamma_{\mu}$ appears more
fundamental. A more general choice would be to take a combination
of  \eqref{1-4b} and \eqref{1-4c} such as
\begin{equation}
\label{1-4d}
\sigma _{\mu \nu} = a_1 \eta _{\mu \nu} +  b_1 \partial _{\mu} \partial _{\nu}~,
\end{equation}
where $a_1$ and $b_1$ are in general Lorentz invariant operators, with the
subscript indicating the rank of the gauge field. In the remainder of
the paper we will take the simple case when $a_1$ and $b_1$ are
constants. Also in most instances when there are two or more options for
formulating a gauge procedure ({\it e.g.} \eqref{1-4b} or \eqref{1-4c}) we
will take the simple case where all the constants but one are set to zero.
For \eqref{1-4d} this means either $a_1=0$ or $b_1=0$.

Next one constructs a ``kinetic" energy term for $A_{\mu}$ by
introducing the field strength tensor
\begin{equation}
\label{1-5}
F^{[1]} _{\mu \nu} = A_1 \partial _{\mu} A_{\nu} + B_1  \partial _{\nu} A_{\mu},
\end{equation}
which is invariant under \eqref{1-4}  \eqref{1-4b} provided the constants
satisfy $A_1+B_1=0$. The standard (Maxwell) theory is obtained by
taking $A_1=-B_1 = 1$.
For the cases \eqref{1-4} \eqref{1-4c}, one would have to impose
some extra condition on the arbitrary $\Gamma_{\mu}$ ({\it e.g.} from \eqref{1-5} $A_1
\partial _{\mu} \Gamma _{\nu} + B_1 \partial _{\nu} \Gamma _{\mu}
=0$) in order to make $F^{[1]} _{\mu \nu}$ invariant. If one does not impose
such a condition on  the arbitrary $\Gamma _{\mu}$,  then one can always make $A_{\mu}
\rightarrow 0$ by  taking $\Gamma_{\mu}=A_{\mu}$. This indicates
for these cases one does not have a dynamical gauge field. We
will later define these cases as ``trivial" or ``semi-trivial".
The reason for writing $F^{[1]} _{\mu \nu}$ in the form given in
\eqref{1-5} is that it will make the connection with the higher
rank field strength tensors more transparent. One can construct
another invariant for just $A_{\mu}$ from $F^{[1]}_{\mu \nu}$
\begin{equation}
\label{1-5a}
G^{[1]} _{\mu} = A_1 \partial _{\mu} \partial ^{\nu} A_{\nu} +
B_1  \partial _{\nu} \partial ^{\nu} A_{\mu}
= \partial ^{\nu} F^{[1]} _{\mu \nu} ~.
\end{equation}
In the standard theory ($A_1=-B_1 = 1$) this is the left hand side
of Maxwell equations with sources. The rank 1 term in \eqref{1-5a}
is mentioned since although it does not play a role in the Maxwell
theory, similar terms occur in the higher rank gauge fields discussed
later, and there they may play some role.
Now taking only  the standard $\sigma _{\mu \nu} =\eta _{\mu \nu}$
case in the construction of $F^{[1]} _{\mu \nu}$ the following
Lagrange density
\begin{equation}
\label{1-6}
{\cal L} _{scalar} = (D ^{[1]}_{\mu} \varphi )^{\ast}
(D ^{[1] \mu} \varphi ) -\frac{1}{4} F ^{[1]} _{\mu \nu} F^{[1] \mu \nu} +... ~~,
\end{equation}
is invariant under the combined  transformations of eq.
(\ref{1-4a}) and the local version of eq. (\ref{1-2}).
The $F^{[1]} _{\mu \nu} F^{[1] \mu\nu}$ term is the
standard Lagrange density of an Abelian U(1)
gauge theory like Maxwell's equations. One can view  Maxwell's
equations as arising from or being derived via the gauge
principle.

The gravitational interaction can be formulated in a somewhat similar
manner.  One can take the global spacetime
symmetries of special relativity and make them local \cite{ramond}
to arrive at a theory of the gravitational interaction. One
again replaces ordinary derivative with covariant derivatives. For
example, the covariant derivative of a vector field, $V_{\nu}$, is
\begin{equation}
\label{1-7}
\partial _{\mu} V_{\nu} \rightarrow \partial _{\mu} V_{\nu} +
\Gamma ^{\alpha} _{\mu \nu} V_{\alpha}~.
\end{equation}
When the metric, $g_{\mu \nu}$, is taken as fundamental then, 
unlike the gauge field for eqs. \eqref{1-3} \eqref{1-4b},
$\Gamma ^{\alpha} _{\mu \nu}$ is not fundamental, but
is defined in terms of the first derivatives of the metric
\begin{equation}
\label{1-8}
\Gamma ^{\alpha} _{\mu \nu} =\frac{1}{2} g^{\alpha \sigma}
(\partial _{\nu} g_{\sigma \mu} + \partial _{\mu} g_{\sigma \nu}
-\partial _{\sigma} g_{\mu \nu})
\end{equation}
In contrast to the standard procedure, the gauge procedure in eq. \eqref{1-4c}
also has the propety that the covariant derivative has the derivatives of
the fundamental quantity, $A_{\mu}$.

\section{Gauge procedure with a rank 0 field}

In this section we present the generalized gauge procedure with a
rank 0 (scalar) gauge field. We begin with the Lagrange density
for a complex scalar matter field of eq. (\ref{1-1}) and a local
phase transformation on this scalar field ({\it i.e.} eq.
\eqref{1-2} with $\Lambda \rightarrow \Lambda (x)$.) The spacetime
dependence of $\Lambda (x)$ means that the derivative of $\varphi$
and $\varphi ^{\ast}$ are no longer invariant under the
transformation in eq. (\ref{1-2}), but become $\partial _{\mu}
\varphi$ $\rightarrow \partial _{\mu} \varphi - i g (\partial
_{\mu} \Lambda) \varphi$ and $\partial _{\mu} \varphi ^{\ast}$
$\rightarrow \partial _{\mu} \varphi ^{\ast} + i g (\partial
_{\mu} \Lambda) \varphi ^{\ast}$ . As in the case of the gauging
procedure with a vector field we want to find a generalization of
the derivative operator, $\partial _{\mu}$, which is invariant
under the local version of eq. (\ref{1-2}). We define this
generalized rank 0 derivative operator as
\begin{equation}
\label{2-2}
D_{\mu} ^{[0]} \equiv \partial _{\mu} - i g \partial _{\mu} \Phi ~.
\end{equation}
$\Phi$ is real, scalar gauge field which is required to undergo the transformation
\begin{equation}
\label{2-3}
\Phi (x) \rightarrow \Phi (x) -  \Lambda (x)
\end{equation}
These transformations of the scalar field are similar to the toy
model considered in \cite{wein}. By replacing the ordinary
derivative in eq. (\ref{1-1}) with  $D _{\mu} ^{[0]}$ the
Lagrangian of eq. (\ref{1-1}) becomes invariant under the local
transformation of eqs. (\ref{1-2}) (\ref{2-3}). As in the
non-standard rank 1 case of \eqref{1-4c} there is no kinetic term,
since from the transformation of \eqref{2-3} it is always possible
to take $\Phi =0$ by choosing $\Lambda = \Phi$. Thus the
Lagrangian
\begin{equation}
\label{2-5} {\cal L} _{scalar} = (D _{\mu} ^{[0]} \varphi )^{\ast}
({D ^{\mu}}^{[0]} \varphi ) = (\partial _{\mu} \varphi ^{\ast} + i
g \varphi ^{\ast}\partial _{\mu} \Phi ) (\partial ^{\mu} \varphi -
i g \varphi \partial ^{\mu} \Phi)
\end{equation}
is invariant under eqs. \eqref{1-2} \eqref{2-3}.
In contrast the standard with covariant derivative, $\partial _{\mu} -ie A_{\mu}$,
the covariant derivative of  \eqref{2-2} involves
the derivative of the fundamental gauge field, $\Phi$. This can
be compared to the covariant derivative of general relativity
which involves derivatives of a more fundamental
quantity: the metric tensor, $g_{\mu \nu}$.

One can apply this rank 0 procedure starting with matter fields
other than a complex scalar field, $\varphi$. In \cite{kato} this
was done starting with a complex, vector field. This was
interpreted as a gauging of the electromagnetic dual symmetry
\cite{jack} \cite{fels} with $\Lambda$ playing the role of the
spacetime dependent rotation angle between electric and magnetic
quantities. In \cite{pak} a related gauging of the Schwarz-Sen
\cite{sen} dual symmetry was given. Aside from gauging the
electromagnetic dual symmetry the idea of having a scalar gauge
field has also been considered by other authors. In ref.
\cite{chaves} an attempt to give a unified version of the Standard
Model was made via the introduction of a generalized covariant
derivative which involved {\it both} vector and scalar gauge
fields.

\section{Classification of gauge procedures}

We have encountered three different gauging procedures: the rank 1
cases of eqs. \eqref{1-4b}, \eqref{1-4c}, and the rank 0 case of
eq. \eqref{2-3}. These illustrate the three general categories of
gauge procedures that we define.
\begin{itemize}
\item {\bf Trivial Case:} This is illustrated by the rank 0 case
of \eqref{2-3}. The gauge field transformation of  \eqref{2-3}
allows one to transform away the gauge field by taking $\Phi =
\Lambda$. This case can be seen as a special case of the standard
gauge procedure with  the association $A_{\mu}  \propto \partial
_{\mu} \Phi$. This is a pure gauge case since for such an
$A_{\mu}$ one finds $F_{\mu \nu} ^{[1]}=0$. For this trivial case
the phase factor and gauge transformation function are related
without the need of a derivative operator (for the rank 0 case the
phase factor and transformation function are the same namely
$\Lambda$). \item {\bf Semi-Trivial Case:} This is illustrated by
the non-standard rank 1 case of \eqref{1-4c}. The transformation
function of the gauge field, $\Gamma_{\mu}$, is arbitrary. By
choosing $A_{\mu} = \Gamma _{\mu}$ it is always possible to
transform away the gauge field, making it a non-dynamical degree
of freedom. Both this and the trivial case have covariant
derivatives of the form
\begin{equation}
\label{2-6}
\partial _{\mu} -i g \partial _{\mu} (Scalar)
\end{equation}
where $Scalar$ is some scalar quantity. The reason for calling
this case semi-trivial is that it has a distinction from the rank
0 case in that the phase factor, $\Lambda$, takes a more restricted
form in terms of the gauge transformation function (the divergence
of $\Gamma_{\mu}$). In contrast to the trivial, the semi-trivial
case phase factor and gauge transformation function are related
using derivative operator(s). \item {\bf  Non-Trivial Case:} This
is illustrated by the rank 1 Maxwell Theory case of \eqref{1-4b}.
In contrast to the previous semi-trivial case, here the phase
factor, $\Lambda$, is arbitrary while the gauge transformation
function, $\Gamma_{\mu}$, takes a restricted form in terms of the
phase function. In this case the gauge field is dynamical and the
covariant derivative does not take the form of \eqref{2-6}.
\end{itemize}

In the following sections we show that for the Abelian phase
symmetries it is possible to formulate a gauge procedure with
gauge fields of various ranks, rather than only
a rank 1, vector field. The procedure will employ generalizations
of \eqref{1-3} \eqref{1-4} of the form
\begin{eqnarray}
\label{2-7}
D_{\mu} &=& \partial _{\mu} -i g \sigma _{\mu _1 \mu _2 ... \mu_{n+1}}
A^{\mu _1 \mu _2 ... \mu_n}  \nonumber \\
A^{\mu _1 \mu _2 ... \mu_n} &\rightarrow & A^{\mu _1 \mu _2 ... \mu_n}
- \Gamma^{\mu _1 \mu _2 ... \mu_n} ~.
\end{eqnarray}
A more general version of \eqref{2-7} would involve linear
combinations of the various $\sigma _{\mu _1 \mu _2 ... \mu_{n+1}}
A^{\mu _1 \mu _2 ... \mu_n}$ operators. As with general relativity
the covariant derivative of these generalized gauge procedures will
sometimes have connections, $\sigma _{\mu _1 \mu _2 ...
\mu_{n+1}}A^{\mu _1 \mu _2 ... \mu_{n-1}}$, which are defined in
terms of the derivative of a more fundamental object. Many of the
theories obtained from this generalized gauging will have
coupling constants with a non-zero mass dimension and
therefore lead to nonrenormalizable
theories as is also the case with general relativity.

There has been previous work on higher rank and higher spin gauge
fields. Rank 2 antisymmetric gauge fields were studied by
Ogievetsky and Palubarinov \cite{ogie}. In an early string theory
paper \cite{kalb} Kalb and Ramond also investigated a rank 2,
antisymmetric gauge field. Fronsdal \cite{fron} and Fronsdal and
Fang \cite{fron1} studied symmetric gauge fields of arbitrary
higher rank. The higher rank gauge fields in this work represented
fields with higher spin. 
Higher rank, anti-symmetric gauge fields were also considered 
in \cite{ohta1} in connection with the U(1) problem in QCD.
More recently Henneaux and Knaepen
\cite{henn} investigated antisymmetric gauge potentials of rank
$>1$. These gauge potentials of rank $n$ naturally couple to
extended objects of dimension $n-1$ and therefore are of interest
for string theory and supergravity which have extended objects. In
\cite{henn} a systematic discussion of the interaction of these
antisymmetric, higher rank gauge fields is given. Finally,
Vasiliev \cite{vasi} \cite{vasi1} has studied a possible
connection of the work in \cite{fron} \cite{fron1} to superstring
theory and the AdS/CFT correspondence. The main difference between
the present discussion of higher rank gauge fields and previous
work is that we begin with the interaction of the higher rank
gauge field {\it i.e.} we first construct the covariant derivative
and then build the gauge invariant field strength tensor. Previous
works focused first on the free, non-interacting gauge field. Also
under the present formulation we find both symmetric and
antisymmetric gauge fields, while previous work had either
antisymmetric or symmetric gauge fields. In the succeeding
sections we will make comparisons and point out the differences
between the present formulation of a gauge principle for higher
rank fields and previous work.

Although in the present paper we do not give a phenomenological
application of the field theories that result from the generalized
gauging procedure, there are several motivations for studying
extensions of the gauge procedure. First, given the central role
played by the gauge principle it is important to investigate
extensions or generalization. Second, some aspects of higher rank
gauge theories discussed here have similarities with general
relativity and may therefore give some insight into a connection
between the gauging of abstract phase symmetries and ``gauging''
of spacetime symmetries. Finally, there are possible applications
for these alternative gauge fields to open questions in cosmology
or hadronic physics. Even some of the ``trivial" or
``semi-trivial" theories discussed in this paper may have physical
applications. As an example, it has been shown \cite{auri}
\cite{hawk} that an antisymmetric, rank 3 Abelian gauge field can
be used to given an account of Einstein's field equation with a
cosmological constant term. In a slightly different
context a discussion of the vacuum energy associated with rank 3 fields can
be found in \cite{ohta2}. In addition ref. \cite{ohta3} 
investigates potentials resulting from gravity coupled
to 3-forms in spirit similar to \cite{auri}.
 
The rank 3 gauge fields in refs.
\cite{auri} \cite{hawk} (see also \cite{auri1}) are non-dynamic
degrees of freedom yet nevertheless they play a physical role. In
\cite{auri2} the rank 3 gauge field was suggested as a possible
candidate for dark matter or dark energy. In \cite{auri3} the rank
3 field was applied to QCD in order to give an analytical
explanation of confinement. By calculating the Wilson loop for
this rank 3 gauge field it was shown that one obtained a static
potential proportional to the volume enclosed ({\it i.e.} $V(R)
\simeq R^3$). Although this is not the standard linear confining
potential ($V(R) \simeq R$) one gets the general picture of
confinement in that it costs an infinite amount of energy to
separate two QCD sources. Since fields like the rank 3 gauge field
of \cite{auri2} \cite{auri3} and the ``trivial" and
``semi-trivial" fields of the present paper are non-dynamical it
is much simpler to obtain analytical results in the path integral
formalism. Although no fundamental scalar field has yet been
experimentally observed, scalar fields have been proposed to
address cosmological ({\it e.g.} dark energy) and particle physics
({\it e.g.} breaking or hiding gauge symmetry) problems. Thus the
non-dynamical scalar field of the previous section may find an
application in this cosmological context as a possible dark energy
candidate. In theories with broken, local symmetries ({\it i.e.}
Landau-Ginzburg theory, or the Higgs mechanism applied to the 
Standard Model of particle physics) 
one encounters scalar fields which develop a vacuum
expectation value. The scalar field of the last section might
find a use in the context of symmetry breaking. Since the scalar
field of the last section is not dynamical one would not expect an
associated propagating scalar particle ({\it e.g.} a Higgs boson).
Thus one might have symmetry breaking without the need for a
propagating scalar particle. We will leave for future work more detailed
studies of the possible applications for both the dynamical and
non-dynamical gauge fields presented in this paper.

\section{Gauge procedure with a rank 2 field}

In this section  we will gauge the local version of the symmetry
of eq. \eqref{1-2} for the Lagrange density in eq. \eqref{1-1}
using a rank 2 gauge field. We define a covariant derivative as
\begin{equation}
\label{4-1}
D ^{[2]} _{\mu} \equiv \partial _{\mu} -
i g \sigma _{\mu \nu \rho}  A^{\nu \rho}~,
\end{equation}
where we have introduced a rank 3 operator, $\sigma _{\mu \nu \rho}$, and a
rank 2 gauge field, $A_{\nu \rho}$. We consider two cases for the symmetry of
the gauge field indices: symmetric and antisymmetric.

\subsection{Symmetric case}

When $A_{\nu \rho}$ is symmetric ($A_{\nu \rho} =A_{\rho \nu}$)
the operator $\sigma _{\mu \nu \rho}$ has the partial symmetry
$\sigma _{\mu \nu \rho} = \sigma _{\mu \rho \nu}$. Constructing
$\sigma _{\mu \nu \rho}$ from $\eta _{\mu \nu}$ and $\partial
_{\mu}$ we can write a general form as
\begin{equation}
\label{4-2}
\sigma _{\mu \nu \rho}= \frac{1}{2} a_2 (\eta _{\mu \nu}
\partial _{\rho} + \eta _{\rho \mu} \partial _{\nu} )
+ b_2 \eta _{\nu \rho} \partial _{\mu} + c_2 \partial _{\mu}
\partial _{\nu} \partial _{\rho}
\end{equation}
where $a_2, b_2, c_2$ are constants with the subscript indicating
the rank of $A_{\mu \nu}$. The last term in \eqref{4-2} has a
greater symmetry (it is symmetric in all three indices) than
required. In conjunction with the transformation of eq.
\eqref{1-2} we require  that $A_{\mu \nu}$ transforms as
\begin{equation}
\label{4-3}
A_{\mu \nu} \rightarrow A_{\mu \nu} - \Gamma_{\mu \nu}.
\end{equation}
If  the rank 2 function $\Gamma _{\mu \nu}$ and $\Lambda$ satisfy
\begin{equation}
\label{4-3a}
\sigma _{\mu \nu \rho} \Gamma^{\nu \rho} -\partial _{\mu} \Lambda =0 ~,
\end{equation}
then the Lagrange density ${\cal L}_{scalar} = (D ^{[2]} _{\mu}
\varphi ) ^{\ast} (D ^{[2] ~\mu} \varphi )+...$ will be invariant
under the combined transformation \eqref{1-2} \eqref{4-3}.  We
will consider three special cases when all but one of the
constants, $a_2, b_2, c_2$, is set to zero
\begin{enumerate}
\item
${\bf [a_2=1, b_2 = c_2=0]}$. In this case the covariant derivative becomes
\begin{equation}
\label{4-3b}
D ^{[2]} _{\mu} = \partial _{\mu} -i g \partial _{\nu} A^{\nu} _{\mu}
\end{equation}
where the symmetry of $A_{\mu \nu}$ was used. The condition
in eq. \eqref{4-3a} becomes
\begin{equation}
\label{4-3c}
\partial _{\nu} \Gamma ^{\nu} _{\mu} - \partial _{\mu} \Lambda =0
\end{equation}
A general solution to \eqref{4-3c} is
\begin{equation}
\label{4-3ca}
\Gamma_{\mu \nu} = \eta _{\mu \nu} \Lambda + h_{\mu \nu} ~,
\end{equation}
where the tensor function satisfies $\partial ^{\nu} h_{\mu \nu} =0$.
From eq. \eqref{4-3c} an example of a specific solution is
\begin{equation}
\label{4-3d}
\Gamma _{\mu \nu} = \partial _{\mu} \partial_ {\nu} f (x) + \eta _{\mu \nu} g(x) ~,
~~~~~~ \Lambda = \partial _{\nu} \partial ^{\nu} f(x) + g(x) + h(x)~,
\end{equation}
where $f(x)$, $g(x)$ and $h(x)$ are arbitrary scalar functions.
The function $h(x)$ must satisfy $\partial _{\mu} h(x) =0$, which
corresponds to the choice $h_{\mu \nu} =
\partial _{\mu} \partial _{\nu} f(x) -
\eta _{\mu \nu} \partial _{\rho} \partial ^{\rho} f(x)$. Thus
${\cal L} _{scalar} = (D ^{[2]} _{\mu} \varphi ) ^{\ast} (D ^{[2]
~\mu} \varphi )+...$, with $D^{[2]} _{\mu}$ defined by
\eqref{4-3b} is invariant under the local phase transformation of
$\varphi (x)$ and the gauge field transformation
\begin{equation}
\label{4-3e}
A_{\mu \nu} \rightarrow A_{\mu \nu} - \partial _{\mu} \partial_{\nu}
f (x) - \eta _{\mu \nu} g(x) ~.
\end{equation}
\item
${\bf [b_2=1, a_2 = c_2=0]}$. In this case the covariant derivative becomes
\begin{equation}
\label{4-3f}
D ^{[2]} _{\mu} = \partial _{\mu} -i g \partial _{\mu} A^{\nu} _{\nu}
\end{equation}
The condition in eq. \eqref{4-3a} becomes
\begin{equation}
\label{4-3g}
\partial _{\mu} \Gamma ^{\nu} _{\nu} - \partial _{\mu} \Lambda =0
\end{equation}
This condition can be satisfied by taking
\begin{equation}
\label{4-3h}
 \Lambda = \Gamma ^{\nu} _{\nu} + f(x) ~,
\end{equation}
with $\partial _{\mu} f(x) =0$. The gauge transformation function,
$\Gamma_{\mu \nu}$,  and phase factor, $ \Lambda$ are arbitrary,
and are related without a derivative operator so this is a trivial
case. Since $\Gamma_{\mu \nu}$ is arbitrary it should always be
possible to transform the gauge field away via $A_{\mu \nu}
\rightarrow A_{\mu \nu} - \Gamma _{\mu \nu}$ with $A_{\mu \nu} =
\Gamma _{\mu \nu}$ making the gauge field non-dynamical. \item
${\bf [c_2=1, a_2 = b_2=0]}$. In this case the covariant derivative
becomes
\begin{equation}
\label{4-3j}
D ^{[2]} _{\mu} = \partial _{\mu} -i g \partial _{\mu} \partial _{\nu}
\partial _{\rho} A^{\nu \rho}
\end{equation}
The condition in eq. \eqref{4-3a} becomes
\begin{equation}
\label{4-3k}
\partial _{\mu} \partial _{\nu} \partial _{\rho}\Gamma ^{\nu \rho} -
\partial _{\mu} \Lambda =0
\end{equation}
This condition can be satisfied by taking
\begin{equation}
\label{4-3l}
 \Lambda = \partial _{\nu} \partial _{\rho}\Gamma ^{\nu \rho} +f(x) ~,
\end{equation}
with $\partial _{\mu} f(x) =0$. This is a semi-trivial case since
the gauge function is arbitrary, but the relationship between it
and the phase factor involves the derivative operator.
\end{enumerate}

For cases 2 and 3  the rank 2 gauge field is arbitrary, and any
form for $\Gamma_{\mu \nu}$ works. For case 1 the specific form
given in \eqref{4-3d} is necessary. This difference can be traced
to the different relationships between the rank 2 gauge field,
$\Gamma_{\mu \nu}$, and the gauge function $\Lambda$ given in eqs.
\eqref{4-3c} \eqref{4-3g} \eqref{4-3k}. Eqs. \eqref{4-3g} and
\eqref{4-3k} involve the same non-contracted indices of the
derivative, while in eq. \eqref{4-3c} the non-contracted indices
of the derivative of the two terms are different. This
is connected with the fact that case 2 is trivial and case 3 is
semi-trivial as previously discussed. Note that cases 2 and 3 have
a covariant derivative of the form of eq. \eqref{2-6}.

Next we want to add a kinetic term involving $A_{\mu \nu}$ alone.
Depending on whether we are dealing with case 1 , 2 or 3 from
above we want to construct a field strength tensor which is
invariant under just the gauge field transformation. Cases 2 and 3
are trivial and semi-trivial so that $\Gamma _{\mu
\nu}$ has a completely arbitrary form. By taking $A_{\mu \nu} =
\Gamma_{\mu \nu}$ it is possible to transform the gauge field
away, therefore it is not a dynamical field. Thus we will only
consider case 1 in constructing the invariant field strength
tensor. Also we will work with the special case of \eqref{4-3d}
where $g(x)=0$. Under these conditions the following rank 3 object
\begin{equation}
\label{4-4}
F ^{[2]} _{\mu \nu \rho} = A_2 \partial _{\mu} A_{\nu
\rho} + B_2 \partial _{\nu} A_{\mu \rho} +C_2 \partial _{\rho}
A_{\mu \nu} ~,
\end{equation}
is invariant under the gauge field transformation of \eqref{4-3d} if the constants
obey $A_2+B_2+C_2=0$. Note that $F ^{[2]} _{\mu \nu \rho}$ is
neither symmetric nor antisymmetric. The defining feature of $F
^{[2]} _{\mu \nu \rho}$ is the permutation of the indices which
generalizes the form of the rank 2 field strength tensor given in
\eqref{1-5}. From $F ^{[2]} _{\mu \nu \rho}$ one can construct
rank 2 and rank 1 objects which are invariant under the gauge
field transformation, for example
\begin{equation}
\label{4-4a}
G^{[2]} _{\mu \nu} = A_2 \partial _{\mu} \partial_{\rho} A^{\rho} _{\nu}
+ B_2 \partial _{\nu} \partial_{\rho} A^{\rho} _{\mu}
+ C_2  \partial _{\rho} \partial^{\rho} A_{\mu \nu} = \partial ^{\rho}
F ^{[2]} _{\mu \nu \rho}
~,~~~~~~~~~~ H^{[2]} _{\mu} = A_2 \partial_{\mu} A^{\rho} _{\rho} +
(B_2+C_2) \partial_{\rho} A^{\rho} _{\mu} = F ^{[2] ~\rho} _{\mu ~ \rho}
\end{equation}
One might exclude $G^{[2]} _{\mu \nu}$ based on the fact that it
involves two derivatives of the gauge field, which means its field
equations would involve three derivatives rather than the standard
two. These are not the most general rank 2 and rank 1 objects. For
rank 2 one could consider terms like $\partial _{\mu} F_{\nu \rho}
^{~\rho}$; for rank 1 one could add terms like $\partial ^{\nu}
\partial ^{\rho} F_{\mu \nu \rho}$.  Keeping the gauge
field-only terms of \eqref{4-4} \eqref{4-4a} one can write a Lagrangian
which is invariant under the local phase and gauge field transformations
\begin{equation}
\label{4-5}
{\cal L} _{scalar} = (D _{\mu} ^{[2]} \varphi )^{\ast}
({D ^{\mu}}^{[2]} \varphi ) + K_1 F^{[2]} _{\mu \nu \rho} F^{[2] \mu \nu \rho} +
K_2 G^{[2]} _{\mu \nu} G^{[2] \mu \nu} + K_3 H^{[2]} _{\mu} H^{[2] \mu} + ....~,
\end{equation}
where $K_1, K_2, K_3$ are constants. The kinetic energy terms
involving only the gauge field, $A_{\mu \nu}$, are more complex
than the rank 1 kinetic terms in  eq. \eqref{1-6}.

For certain choices of the constants one can find Maxwell-like
Lagrangian embedded in \eqref{4-5}. One example is to take
$K_1=-\frac{1}{4} , ~K_2=K_3=0$ and $A_2=-B_2=1 , ~C_2=0$ reducing
\eqref{4-5}
\begin{equation}
\label{4-6}
{\cal L} _{scalar} = (D _{\mu} ^{[2]} \varphi )^{\ast}
({D ^{\mu}}^{[2]} \varphi ) -\frac{1}{4}
(\partial _{\mu} A_{\nu \rho} - \partial _{\nu} A_{\mu \rho})
(\partial ^{\mu} A^{\nu \rho} - \partial ^{\nu} A^{\mu \rho}) + ....
\end{equation}
The pure gauge field term above looks like that of the rank 1
Maxwell case with $A_{\mu} \rightarrow A_{\mu \rho}$. Another
example is to take $K_2=-\frac{1}{4} , ~K_1=K_3=0$ and $A_2=-B_2=1
, ~C_2=0$ reducing \eqref{4-5}
\begin{equation}
\label{4-7}
{\cal L} _{scalar} = (D _{\mu} ^{[2]} \varphi )^{\ast}
({D ^{\mu}}^{[2]} \varphi ) -\frac{1}{4}
(\partial _{\mu} \partial _{\rho} A_{\nu} ^{\rho} -
\partial _{\nu} \partial _{\rho} A_{\mu} ^{\rho})
(\partial ^{\mu} \partial _{\rho} A^{\nu \rho} - \partial ^{\nu}
\partial _{\rho}A^{\mu \rho}) + ....
\end{equation}
Here the pure gauge field term looks like the rank 1 Maxwell case
but with $A_{\mu} \rightarrow
\partial _{\rho} A_{\mu} ^{\rho}$.

The mass dimension of $g$ ($M[g]$) indicates whether
the theory is renormalizable or not: for $M[g] \ge 0$ the theory
is naively renormalizable , while for $M[g]<0$ the theory is
nonrenormalizable. For $g$ the mass dimension can be determined
from eq. \eqref{4-1}, and noting that the derivative operator has
mass dimension $+1$ and taking the gauge field, $A_{\mu \nu}$, to
have the usual mass dimension $+1$. For case
1 from \eqref{4-3b} the coupling $g$ has
mass dimension $-1$, implying that the
Lagrangian in eq. (\ref{4-5}) is nonrenormalizable.

\subsection{Antisymmetric case}

When $A_{\nu \rho}$ is antisymmetric ($A_{\nu \rho} =-A_{\rho
\nu}$) the operator $\sigma _{\mu \nu \rho}$ has the partial
antisymmetry $\sigma _{\mu \nu \rho} = -\sigma _{\mu \rho \nu}$.
In addition to $\eta _{\mu \nu}$ and $\partial _{\mu}$ we now
consider $\epsilon _{\mu \nu \rho \sigma}$  (the totally
antisymmetric 4d tensor) in constructing $\sigma _{\mu \nu \rho}$.
A general form is
\begin{equation}
\label{4-3n}
\sigma _{\mu \nu \rho} = \frac{1}{2} a_2 (\eta _{\mu \nu}
\partial _{\rho} - \eta _{\rho \mu} \partial _{\nu} )
+ b_2 \epsilon _{\sigma \mu \nu \rho} \partial ^{\sigma} .
\end{equation}
$a_2 , b_2$ are constants. The last term has a greater degree of
antisymmetry than required of  $\sigma _{\mu \nu \rho}$. We
consider the special cases when one or the other of these
constants is set to zero.
\begin{enumerate}
\item ${\bf [a_2=1, b_2 =0]}$. If we take $\Gamma_{\mu \nu}$ to be
antisymmetric as $A_{\mu \nu}$ then from \eqref{4-3a} the
relationship between $\Gamma_{\mu \nu}$ and $\Lambda$ is
\begin{equation}
\label{4-3o}
\partial _{\nu} \Gamma ^{\nu} _{\mu} - \partial _{\mu} \Lambda =0 ~,
\end{equation}
which is the same as \eqref{4-3c}. However we can not take the same solution as in
\eqref{4-3d}, since this makes $\Gamma_{\nu \mu}$ symmetric. We have not found a
solution which satisfies \eqref{4-3o} and has an antisymmetric $\Gamma_{\mu \nu}$.
\item
${\bf [b_2=1, a_2 =0]}$.
From \eqref{4-3a} the relationship between $\Gamma_{\mu \nu}$ and $\Lambda$ is
\begin{equation}
\label{4-3p} \epsilon _{\sigma \mu \nu \rho} \partial ^{\sigma}
\Gamma ^{\nu \rho} - \partial _{\mu} \Lambda =0 ~.
\end{equation}
We have not found a solution to this relationship as in the symmetric cases.
\end{enumerate}
In both antisymmetric cases we have not found a $\Gamma_{\mu \nu}$
which satisfies either \eqref{4-3o} or \eqref{4-3p}, and the
correct symmetry of $\Gamma_{\mu \nu}$. Without such a
relationship it is not possible to construct a kinetic energy term
for the antisymmetric $A_{\mu \nu}$.

As mentioned in the previous section other authors \cite {ogie,
kalb, fron, fron1} have considered rank 2 gauge fields. In ref.
\cite{kalb} the gauge fields considered were antisymmetric, rank 2
gauge fields. The field strength tensor ({\it i.e.} $F_{\mu \nu
\rho} = \partial _{\mu} A_{\nu \rho} + \partial _{\nu} A_{\rho
\mu}+\partial _{\rho} A_{\mu \nu}$) and the gauge transformation
({\it i.e.} $A_{\mu \nu} \rightarrow A_{\mu \nu} + \partial _{\mu}
\Lambda _{\nu} - \partial _{\nu} \Lambda _{\mu}$) of \cite{kalb} can be seen to
resemble those of the present work in \eqref{4-4} \eqref{4-3e},
differing by having an antisymmetric rather than symmetric
character. There is a closer similarity between the equations of
\eqref{4-4} \eqref{4-3e}, and
those in \cite{fron, fron1} where symmetric, rank 2 gauge fields
are given. The main distinction between the present higher rank
gauge fields and previous studies is that here we begin
constructing the gauge field from its coupling to some matter
field, and then construct the field strength tensor. In
the work of \cite{kalb, fron, fron1} the gauge field, its
transformation, and the field strength tensor are constructed
first without coupling the gauge field to some matter field. 

\section{Gauge procedure with a rank 3 and higher fields}

In this section  we will gauge the local version of the symmetry
of eq. \eqref{1-2} for the Lagrange density in eq. \eqref{1-1}
using a rank 3 gauge field. From this example it should become
clear how to gauge symmetries like \eqref{1-2} with rank $n$
fields. As $n$ becomes large the number of possible terms in the
definition of $\sigma _{\mu _1 \mu_2 ....\mu_{n+1}}$ and in the
construction of a kinetic term for the gauge field becomes larger
and more complex. We define a covariant derivative as
\begin{equation}
\label{5-1}
D ^{[3]} _{\mu} \equiv \partial _{\mu} - i g \sigma _{\mu \nu \rho \tau}
A^{\nu \rho \tau}~,
\end{equation}
where we have introduced a rank 4 operator, $\sigma _{\mu \nu \rho \tau}$, and a
rank 3 gauge field, $A_{\nu \rho \tau}$. We will consider the two cases when
this rank 3 gauge field is either totally symmetric or totally antisymmetric in its
indices. Cases when there is a mixed symmetry of the indices will be considered
in a future work.

\subsection{Symmetric case}

If $A_{\nu \rho \tau}$ is taken to be totally symmetric then
$\sigma_{\mu \nu \rho \tau}$ must be symmetric in its last three
indices. Using the operators, $\eta _{\mu \nu}$ and $\partial
_{\mu}$ we can write a general form for $\sigma _{\mu \nu \rho
\tau}$ which is symmetric in the last three indices
\begin{eqnarray}
\label{5-2}
\sigma _{\mu \nu \rho \tau} &=&
\frac{1}{3} a_3 (\eta _{\mu \nu} \eta_{\rho \tau} + \eta _{\mu \rho} \eta_{\nu \tau}
+\eta _{\mu \tau} \eta_{\rho \nu})
+ \frac{1}{3}b_3( \eta _{\mu \nu} \partial _{\rho}
\partial _{\tau} +\eta _{\mu \rho} \partial _{\nu} \partial _{\tau}
+\eta _{\mu \tau} \partial _{\rho} \partial _{\nu}) \nonumber \\
&+& \frac{1}{3}c_3( \eta _{\rho \nu}
\partial _{\mu} \partial _{\tau} +\eta _{\tau \rho} \partial _{\mu} \partial _{\nu}
+\eta _{\nu \tau} \partial _{\mu} \partial _{\rho})
+ d_3 \partial _{\mu} \partial _{\nu} \partial _{\rho} \partial _{\tau}
\end{eqnarray}
where $a_3, b_3, c_3, d_3$ are constants with the subscript
indicating the rank of $A_{\mu \nu \tau}$. The last term in
\eqref{5-2} has a greater index symmetry than required: it is
symmetric in all four indices. In conjunction with the
transformation of eq. \eqref{1-2} we now require  $A_{\mu \nu\tau}$
to transform as
\begin{equation}
\label{5-3}
A_{\mu \nu \tau} \rightarrow A_{\mu \nu \tau} - \Gamma_{\mu \nu \tau} ~,
\end{equation}
with  the rank 3 function $\Gamma _{\mu \nu \tau}$ and $\Lambda$ satisfying
\begin{equation}
\label{5-3a}
\sigma _{\mu \nu \rho \tau} \Gamma ^{\nu \rho \tau} -\partial _{\mu} \Lambda =0 ~,
\end{equation}
then the Lagrange density ${\cal L}_{scalar} = (D ^{[3]} _{\mu} \varphi ) ^{\ast}
(D ^{[3] ~\mu} \varphi )+...$ will be invariant under the combined transformation
\eqref{1-2} \eqref{5-3}.  As in the previous section we will consider three special
cases when all but one of the constants, $a_3, b_3, c_3, d_3$, are set to zero
\begin{enumerate}
\item
${\bf [a_3=1, b_3 = c_3=d_3=0]}$. In this case the covariant derivative becomes
\begin{equation}
\label{5-3b}
D ^{[3]} _{\mu} = \partial _{\mu} -i g A _{\mu \rho} \;^{\rho}
\end{equation}
where the symmetry of $A_{\mu \nu \rho}$ was used. The condition in
eq. \eqref{5-3a} becomes
\begin{equation}
\label{5-3c}
\Gamma _{\mu \rho} \;^{\rho} - \partial _{\mu} \Lambda =0
\end{equation}
An example of a solution to \eqref{5-3c} is
\begin{equation}
\label{5-3d} \Gamma _{\mu \nu \rho} = \partial _{\mu} \partial_
{\nu} \partial _{\rho} f (x) + \frac{1}{6} (\eta _{\mu \nu}
\partial _{\rho} + \eta_{\nu \rho} \partial_{\mu} + \eta_{\rho
\mu} \partial_{\nu}) g(x) ~, ~~~~~~ \Lambda = \partial _{\rho}
\partial ^{\rho} f(x) + g(x) + h(x)~,
\end{equation}
where $f(x), g(x), h(x)$ are arbitrary functions with $\partial
_{\mu} h(x) =0$. Eq. \eqref{5-3d} can be seen as a generalization
of  the rank 2 example solution of \eqref{4-3d}. Thus ${\cal L}
_{scalar} = (D ^{[3]} _{\mu} \varphi ) ^{\ast} (D ^{[3] ~\mu}
\varphi )+...$, with $D^{[3]} _{\mu}$ defined by \eqref{5-3b} is
invariant under the phase transformation of  $\varphi (x)$ and the
gauge field transformation
\begin{equation}
\label{5-3e} A_{\mu \nu \rho} \rightarrow A_{\mu \nu \rho} -
\partial _{\mu} \partial_{\nu} \partial _{\rho} f (x) -\frac{1}{6}
(\eta _{\mu \nu} \partial _{\rho} + \eta_{\nu \rho} \partial_{\mu}
+ \eta_{\rho \mu} \partial_{\nu}) g(x)  ~.
\end{equation}
This gauge field transformation appears as a generalized version
of the rank 2 gauge field case given in \eqref{4-3e}. This case is
closer to the standard vector gauge procedure of \eqref{1-4b} than
any of the previous rank 2 cases, since the  covariant derivative
in \eqref{5-3b} only involves the gauge field and not derivatives
of the gauge field.
\item ${\bf [b_3=1, a_3 = c_3=d_3=0]}$. In
this case the covariant derivative becomes
\begin{equation}
\label{5-3f}
D ^{[3]} _{\mu} = \partial _{\mu} -i g \partial ^{\rho}
\partial^{\tau} A_{\mu \rho \tau}
\end{equation}
The condition in eq. \eqref{5-3a} becomes
\begin{equation}
\label{5-3g}
\partial ^{\nu} \partial^{\rho} \Gamma_{\mu \nu \rho} - \partial _{\mu} \Lambda =0~.
\end{equation}
Equation \eqref{5-3g} can be satisfied by taking the same $\Gamma
_{\mu \nu \rho}$  as in to \eqref{5-3d}, and a phase factor of the
form
\begin{equation}
\label{5-3h}
\Lambda = \partial _{\rho} \partial ^{\rho} \partial _{\tau} \partial ^{\tau} f(x)
+ \partial _{\tau} \partial ^{\tau} g(x) + h(x) ~,
\end{equation}
where $f(x), g(x), h(x)$ are arbitrary functions with $\partial
_{\mu} h(x) =0$. Although the gauge transformations of $A_{\mu \nu
\rho}$ are identical for the two cases, the phase factor,
$\Lambda$, now involves  two d'Alembertian operators, $\partial
_{\mu} \partial ^{\mu}$ instead of one.
\item ${\bf [c_3=1, a_3 =b_3=d_3=0]}$. In this case the covariant
derivative becomes
\begin{equation}
\label{5-3fa}
D ^{[3]} _{\mu} = \partial _{\mu} -i g \partial _{\mu} \partial^{\tau}
A ^{\rho} _{\rho \tau}
\end{equation}
The condition in eq. \eqref{5-3a} becomes
\begin{equation}
\label{5-3ga}
\partial _{\mu} \partial^{\tau} \Gamma ^{\rho} _{\rho \tau} -
\partial _{\mu} \Lambda =0
\end{equation}
This condition can be satisfied by taking
\begin{equation}
\label{5-3ha}
 \Lambda =  \partial ^{\tau} \Gamma_{\rho \tau} ^{\rho} + f(x)~,
\end{equation}
where $\partial _{\mu} f =0$. Since the covariant derivative is of
the form \eqref{2-6} and the gauge function and phase factor are
related using derivatives this case is characterized as
semi-trivial.  The Lagrangian ${\cal L} _{scalar} = (D ^{[3]}
_{\mu} \varphi ) ^{\ast} (D ^{[3] ~\mu} \varphi )+...$, with
$D^{[3]} _{\mu}$ defined by \eqref{5-3fa} is invariant under the
phase transformation of  $\varphi (x)$ and an arbitrary gauge
transformation $A_{\mu \nu \rho} \rightarrow A_{\mu \nu \rho} -
\Gamma_{\mu \nu \rho}$.
\item ${\bf [d_3=1, a_3 = b_3=c_3=0]}$. In
this case the covariant derivative becomes
\begin{equation}
\label{5-3j}
D ^{[3]} _{\mu} = \partial _{\mu} -i g \partial _{\mu} \partial _{\nu}
\partial _{\rho} \partial _{\tau} A^{\nu \rho \tau}
\end{equation}
The condition in eq. \eqref{5-3a} becomes
\begin{equation}
\label{5-3k}
\partial _{\mu} \partial _{\nu} \partial _{\rho} \partial _{\tau}
\Gamma^{\nu \rho \tau}
 - \partial _{\mu} \Lambda =0
\end{equation}
This condition can be satisfied by taking
\begin{equation}
\label{5-3l}
 \Lambda = \partial _{\nu} \partial _{\rho} \partial _{\tau} \Gamma^{\nu \rho \tau}
 + f(x) ~,
\end{equation}
where $\partial _{\mu} f =0$. As for the preceding case this is also semi-trivial.
\end{enumerate}

As before we want to add a kinetic term involving  $A_{\mu \nu
\rho}$ alone. For cases 1 and 2 the transformation of $A_{\mu \nu
\rho}$ is the same and is given by \eqref{5-3e}. Thus both cases
will have the same pure gauge, invariant field strength tensors.
Both cases 3 and 4 are semi-trivial and have completely arbitrary
forms for the gauge transformation. In constructing the
invariant field strength tensor we only consider the non-trivial
cases 1 and 2. The rank 4 object
\begin{equation}
\label{5-4}
F ^{[3]} _{\mu \nu \rho \tau} = A_3 \partial _{\mu} A_{\nu \rho \tau}
+ B_3 \partial _{\nu} A_{\rho \tau \mu}
+C_3 \partial _{\rho} A_{\tau \mu \nu} + D_3 \partial _{\tau} A_{\mu \nu \rho}~,
\end{equation}
is invariant under \eqref{5-3e} if the constants obey
$A_3+B_3+C_3+D_3=0$. The common feature between this invariant
field strength tensor and the rank 1 and 2 cases of \eqref{1-5}
and \eqref{4-4} is the permutation of indices. From $F ^{[3]}
_{\mu \nu \rho \tau}$ one can construct lower rank invariants
which may play some physical role. Some examples of invariant rank
3, 2 and 1 objects are
\begin{equation}
\label{5-4a}
G^{[3]} _{\mu \nu \rho} = \partial ^{\tau} F ^{[3]} _{\mu \nu \rho \tau} ,
~~~~~~~~~
H^{[3]} _{\mu \nu} = F ^{[3] ~\rho} _{\mu \nu \rho} ,
~~~~~~~~~
J^{[3]} _{\mu} = \partial ^{\nu} F ^{[3] ~\rho} _{\mu \nu \rho} ~,
\end{equation}
Combining all these results gives
\begin{equation}
\label{5-5} {\cal L} _{scalar} = (D _{\mu} ^{[3]} \varphi )^{\ast}
({D ^{\mu}}^{[3]} \varphi ) + K_1 F^{[3]} _{\mu \nu \rho\tau}
F^{[3] \mu \nu \rho\tau} + K_2 G^{[3]} _{\mu \nu \rho} G^{[3] \mu
\nu \rho} + K_3 H^{[3]} _{\mu \nu} H^{[3] \mu \nu} + K_4 J^{[3]}
_{\mu} J^{[3] \mu} + ....
\end{equation}
which is invariant under the local phase transformation and the
gauge field transformation eq. \eqref{5-3e}. The $K_i$'s are
constants. One could reduce the complexity of the pure gauge terms
by making assumptions or restrictions. For example one could drop
the $G^{[3] \mu \nu \rho}$ and $J^{[3] \mu}$ terms based on the
fact that they have more than one derivative operator acting on
the gauge field. This implies that the field equations for $A_{\mu
\nu \rho}$ coming from these terms would be higher than second
order in the derivatives. As in the rank 2 case one can find
Maxwell-like equations embedded in the above Lagrangian. One
simple example is to set $K_1=-\frac{1}{4} , ~K_2=K_3=K_4=0$ and
$A_3=-B_3=1 , ~C_3=D_3=0$ in \eqref{5-4}. Then the Lagrange
density in eq. \eqref{5-5} becomes
\begin{equation}
\label{5-6}
{\cal L} _{scalar} = (D _{\mu} ^{[3]} \varphi )^{\ast}
({D ^{\mu}}^{[3]} \varphi ) -\frac{1}{4} (\partial _{\mu} A_{\nu \rho \tau}
- \partial _{\nu} A_{\mu \rho \tau})
(\partial ^{\mu} A^{\nu \rho \tau} - \partial ^{\nu} A^{\mu \rho \tau}) + ....
\end{equation}
The second term looks like that of the Maxwell kinetic term with $A_{\mu} \rightarrow
A_{\mu \nu \rho}$

For the symmetric rank 3 gauge field the dimension of the coupling
and therefore whether the theory is renormalizable or not is
different for each of the three cases above. For case $1$, there
are no derivative terms which appear in the second term in the
covariant derivative in \eqref{5-3b}. Taking $A_{\mu \nu
\rho}$ to have the usual mass dimension of $+1$, then $g$ is
dimensionless. Thus this case may lead to a renormalizable theory.
For case $2$ two derivatives appear in the second term of the
covariant derivative in \eqref{5-3f}. Therefore taking $A_{\mu \nu
\rho}$ to have mass dimension of $+1$, implies that $g$ should
have mass dimension $-2$. Finally, for case $3$ four derivatives
appear in the second term of the covariant derivative in
\eqref{5-3j}. Again taking $A_{\mu \nu \rho}$ to have mass
dimension of $+1$, gives $g$ a mass dimension of $-4$.

\subsection{Antisymmetric case}

For the case when $A_{\nu \rho \tau}$ is antisymmetric in all its
indices the operator $\sigma _{\mu \nu \rho \tau}$ should be
antisymmetric in its last three indices. Using $\eta _{\nu \nu}$ ,
$\partial _{\mu}$, and $\epsilon _{\mu \nu \rho \tau}$ to
construct $\sigma _{\mu \nu \rho \tau}$ we find that only the
simple case
\begin{equation}
\label{5-8}
\sigma _{\mu \nu \rho \tau} =  \epsilon _{\mu \nu \rho \tau}
\end{equation}
satisfies the antisymmetry. Actually, eq. \eqref{5-8} satisfies a
greater antisymmetry than required since $\epsilon _{\mu \nu \rho
\tau}$ is antisymmetric in all four indices. In this case the
covariant derivative becomes \begin{equation} \label{5-8a} D
^{[3]} _{\mu} = \partial _{\mu} -i g \epsilon _{\mu \nu \rho \tau}
A^{\nu \rho \tau}
\end{equation}
With \eqref{5-8} the condition in eq. \eqref{5-3a}
becomes
\begin{equation}
\label{5-8b}
\epsilon_{\mu \nu \rho \tau} \Gamma^{\nu \rho \tau} - \partial _{\mu} \Lambda =0
\end{equation}
This condition can be satisfied by
\begin{equation}
\label{5-8d}
\Gamma ^{\nu \rho \tau} = -\frac{1}{6} \epsilon ^{\alpha \nu \rho \tau}
\partial _{\alpha} \Lambda ~,
\end{equation}
where the identity $\epsilon ^{\alpha \nu \rho \tau} \epsilon
_{\mu \nu \rho \tau}= - 6 \delta ^{\alpha} _{\mu}$ was used. Note
the difference from the previous symmetric cases in eqs.
\eqref{5-3d}, \eqref{5-3h} where both $\Gamma_{\nu \rho \tau}$ and
$\Lambda (x)$ are defined in terms of the derivatives of some
auxiliary function, $f(x)$, or in eq. \eqref{5-3l} where $\Lambda
(x)$ is defined in terms of the derivatives of $\Gamma_{\nu \rho
\tau}$. In eq. \eqref{5-8d} $\Gamma_{\nu \rho \tau}$ is defined
directly in terms of the derivative of $\Lambda$. Except for the
factor of $-\frac{1}{6} \epsilon ^{\alpha \nu \rho \tau}$ this is
similar to the standard, rank 1 case of \eqref{1-4b}. Thus ${\cal
L} _{scalar} = (D ^{[3]} _{\mu} \varphi ) ^{\ast} (D ^{[3] ~\mu}
\varphi )+...$, with $D^{[3]} _{\mu}$ defined by \eqref{5-8a} is
invariant under
\begin{equation}
\label{5-8e}
\varphi (x) \rightarrow e^{-ig \Lambda (x)}
\varphi (x) ~,~~~~~~
\varphi ^{\ast} (x) \rightarrow e^{ig \Lambda (x) }
\varphi ^{\ast} (x) ~, ~~~~~~~
A_{\nu \rho \tau} \rightarrow A_{\nu \rho\tau} + \frac{1}{6}
\epsilon _{\alpha \nu \rho \tau} \partial ^{\alpha} \Lambda ~.
\end{equation}
The gauge transformation for $A_{\nu \rho \tau}$ is similar to
that for the rank 3, antisymmetric gauge field discussed in
\cite{auri}, \cite{hawk} , \cite{auri3}. As in the symmetric case
we want to construct a field strength tensor which is invariant
under just the last transformation in \eqref{5-8e}, and from which
a gauge field kinetic energy term can be built. Due to the
antisymmetric nature of $A_{\mu \nu \rho}$ we have not been able
to find  rank 4, rank 3, or rank 1 invariant terms like those for
the symmetric case in eqs. \eqref{5-4} or \eqref{5-4a}. The
following rank 2 term is invariant
\begin{equation}
\label{5-9}
{\cal H}^{[3]} _{\mu \nu} = \partial ^{\tau} A_{\tau \mu \nu} =
\eta ^{\lambda \tau} \partial _{\lambda} A_{\tau \mu \nu}
\end{equation}
One may worry that it is always possible to choose a Lorentz like
gauge condition ($\partial _{\tau} A^{\tau \mu \nu} =0$) so that
${\cal H}^{[3]} _{\mu \nu}=0$. However the transformation for
$A_{\nu \mu \rho}$ given in  \eqref{5-8e} is not the standard
gauge transformation, and if the $\partial ^{\tau} A_{\tau \mu
\nu} \ne 0$ to begin with the gauge transformation can not be used
to make it so since $\epsilon _{\alpha \nu \rho \tau} \partial
^{\tau} \partial ^{\alpha} \Lambda =0$. Combining eqs.
\eqref{5-8a} and \eqref{5-9} gives
\begin{equation}
\label{5-10}
{\cal L} _{scalar} = (D _{\mu} ^{[3]} \varphi )^{\ast}
({D ^{\mu}}^{[3]} \varphi ) + K_1 {\cal H}^{[3]} _{\mu \nu}
{\cal H}^{[3] \mu \nu}+ ....
\end{equation}
which is invariant under the local transformation eq.
\eqref{5-8e}. $K_1$ is an arbitrary constant. We calculate the
field equations for $A_{\mu \nu \rho}$ coming from \eqref{5-10} in
steps. First
\begin{eqnarray}
\label{5-11}
\partial _{\lambda} \frac{\partial {\cal L} _{scalar}}{\partial (\partial _{\lambda}
A_{\tau \mu \nu})}
&=&
2 K_1 \partial _{\lambda} \left( \eta ^{\lambda \tau}
\eta _{\sigma \rho} \partial ^{\sigma}
A^{\rho \mu \nu} \right) =
2 K_1 \left( \partial ^{\tau} \partial _{\rho} A^{\rho \mu \nu} \right)
\nonumber \\
\frac{\partial {\cal L} _{scalar}}{\partial  A_{\tau \mu \nu}} &=&
-12 g^2 \varphi ^{\ast} \varphi A^{\tau \mu \nu}
+ i g \epsilon ^{\rho \tau \mu \nu} \left( \varphi ^{\ast} \partial _{\rho} \varphi
- \varphi \partial _{\rho} \varphi ^{\ast} \right)
\end{eqnarray}
In calculating the $-12 g^2 \varphi ^{\ast} \varphi A_{\alpha
\beta \gamma}$ part of the second line the identity
\begin{equation}
\label{5-12}
-\epsilon ^{\mu \alpha \beta \gamma} \epsilon _{\mu \nu \rho \sigma} =
\delta ^{\alpha} _{\nu} ( \delta ^{\beta} _{\rho} \delta ^{\gamma} _{\sigma} -
\delta ^{\beta} _{\sigma} \delta ^{\gamma} _{\rho})
-\delta ^{\alpha} _{\rho} ( \delta ^{\beta} _{\nu} \delta ^{\gamma} _{\sigma} -
\delta ^{\beta} _{\sigma} \delta ^{\gamma} _{\nu})
+\delta ^{\alpha} _{\sigma} ( \delta ^{\beta} _{\nu} \delta ^{\gamma} _{\rho} -
\delta ^{\beta} _{\rho} \delta ^{\gamma} _{\nu}) ~,
\end{equation}
was used. Putting this all together gives the following equation
for the antisymmetric, rank 3 gauge field
\begin{equation}
\label{5-13}
2 K_1 \left( \partial ^{\tau} \partial _{\rho} A^{\rho \mu \nu} \right) =
-12 g^2 \varphi ^{\ast} \varphi A^{\tau \mu \nu}
+ i g \epsilon ^{\rho \tau \mu \nu} \left( \varphi ^{\ast} \partial _{\rho} \varphi
- \varphi \partial _{\rho} \varphi ^{\ast} \right)~.
\end{equation}
This is not the standard wave equation that is usually encountered
for gauge fields. Rather than the d'Alembertian, $\partial _{\mu}
\partial ^{\mu}$, one finds the 4-gradient of the 4-divergence of
$A_{\mu \nu \rho}$. In the present paper we do not discuss any
possible physical use of this rank 3, antisymmetric case. However,
from \eqref{5-8a} and \eqref{5-8e} the coupling $g$ is
dimensionless, implying that this unusual theory should be
renormalizable.

\section{Summary and Conclusion}

In this paper we have presented various gauging procedures for a
phase symmetry using gauge fields having ranks other than 1.  Here
we summarize the overall structure of this gauging method:
\begin{enumerate}
\item
Starting from some initial matter field (such as the complex
scalar field, $\varphi$, used throughout this paper) one imposes a
local version of the phase symmetry in eq. \eqref{1-2}. One then
introduces a covariant derivative of the form of the first
equation in \eqref{2-7} and a rank $n$ gauge field which
transforms like the second equation in  \eqref{2-7} \item The
definition of the covariant derivative in \eqref{2-7} involves the
introduction of a rank $n+1$ operator, $\sigma _{\mu_1 \mu _2 ....
\mu _{n+1}}$, which is constructed from the derivative operator,
$\partial _{\mu}$, and the metric tensor, $\eta _{\mu \nu}$, and
(when the gauge field is antisymmetric) $\epsilon_{\mu \nu \rho
\sigma}$ . There is some freedom in the construction of $\sigma
_{\mu_1 \mu _2 .... \mu _{n+1}}$ as can be seen from \eqref{1-4b},
\eqref{1-4c}, \eqref{4-2}, \eqref{4-3n},  \eqref{5-2}, and
\eqref{5-8}. \item There were three different categories of gauge
procedures: trivial, semi-trivial, or non-trivial. The trivial
cases ({\it e.g.} the rank 0 case of \eqref{2-3}) involved only
the introduction of an arbitrary phase factor, $\Lambda$, in terms
of which one could define the gauge function, $\Gamma _{\mu _1
\mu _2 ...}$, without the use of the derivative operator. The
semi-trivial case ({\it e.g.} the rank 1 case of \eqref{1-4c})
involved the introduction of an arbitrary  gauge function,
$\Gamma_{\mu _1 \mu _2 ....}$, in terms of which the phase factor
was defined. For the semi-trivial case the relationship between
the phase factor and the gauge function involved the derivative
operator. The non-trivial case  ({\it e.g.} the rank 1 case of
\eqref{1-4b}) involved the introduction of an arbitrary phase
factor, $\Lambda$, in terms of which the gauge function was
defined. For the trivial and semi-trivial cases the gauge field
could always be transformed away and thus was not dynamical. For
the non-trivial case one could construct an invariant field
strength tensor and a kinetic energy term. \item For the rank $n$,
symmetric gauge field one can define a rank $n+1$ field strength
tensor (as in eqs. \eqref{1-5}, \eqref{4-4}, \eqref{5-4}) which
was invariant under just the transformation of the gauge field,
eqs. \eqref{4-3e} or \eqref{5-3e}. This allows the construction of
kinetic terms for the gauge field in the Lagrangian. From these
rank $n+1$ invariants one can construct a host of lower rank
invariants. Although the rank $n+1$ field strength tensor has a
permutation symmetry among its indices there is a large degree of
arbitrariness in the construction of the lower rank, gauge
field-only terms. By making special choices it is possible to have
the Lagrangian for the rank $n$ symmetric gauge field take a
Maxwell-like form. Without these special choices one obtains more
complicated field equations for the gauge fields. For the
antisymmetric cases the situation with respect to constructing a
field strength tensor is not as clear as in the symmetric case.
For the rank 2 case we were not able to find a proper field
strength tensor, and for the rank 3 case we found a field strength
tensor which yielded non-standard field equations. \item The
coupling constant, $g$, in general will have a non-zero mass
dimension which can be determined by the covariant derivative, the
structure of the $\sigma _{\mu _1 \mu _2 ... \mu_{n+1}}$ and
taking the gauge field, $A_{\mu _1 \mu _2 .... \mu _m}$, to have
mass dimension $+1$. The mass dimension of $g$ is then the inverse
of $\sigma _{\mu _1 \mu _2 ... \mu_{n+1}}$. For example,
\eqref{4-3b} and \eqref{4-3e} imply $g$ has a mass dimension of
$-1$; eqs. \eqref{1-4c} and \eqref{5-3e} imply $g$ has a mass
dimension of $-2$. These theories having a dimensionful coupling
are nonrenormalizable. There are cases ({\it e.g.} the standard
rank 1 of \eqref{1-4b} case or the rank 3 case of \eqref{5-3b})
where $g$ has mass dimension $0$ and should therefore be
renormalizable.
\end{enumerate}
There have been other studies of higher rank ({\it i.e.} higher
spin) gauge fields to which the present work can be compared and
contrasted. In particular the work of Fronsdal has sought to
extend a gauge procedure to higher rank fields of both integer
\cite{fron} and half integer spin \cite{fron1}. In \cite{fron,
fron1} the gauge transformation of the gauge fields is somewhat
different from those in the present work. In \cite{fron} the
transformation of the rank $n$ gauge field involves one derivative
operator acting on rank $n-1$ gauge parameters. Here the
transformation of the rank $n$ gauge field involves $n$ derivative
operators acting on a rank 0 gauge parameter (see eqs.
\eqref{1-4}, \eqref{2-3} , \eqref{4-3} and \eqref{5-3}).  In both
the present work and in \cite{fron} the rank $n \ge 2$ gauge
fields are totally symmetric under exchange of indices. However,
in \cite{fron} the gauge fields must satisfy a traceless
condition, unlike the gauge fields in the present work.
A current review of these higher spin gauge theories, and possible
applications to supersymmetry and string theory can be found in
\cite{vasi}. The biggest distinction between the present higher
rank gauge theory and the previous work (either \cite{fron, fron1}
or also \cite{kalb}) is that here that gauge procedure is
developed by starting with the coupling to some matter field and
then constructing the gauge field transformation and invariant
field strength tensor from this point. In \cite{kalb, fron1} the
gauge transformation and invariant field strength tensor is
constructed without coupling the gauge field to any matter field.

We have not attempted to give a physical application for the
generalized gauging procedure in this paper. Given the importance
of the gauge principle to modern field theory it is useful to
explore generalizations which may indicate a larger structure, and
can show connections between the local abstract symmetries of
particle physics, and the local spacetime symmetries of general
relativity. In future work we will investigate possible
cosmological and hadronic applications of the rank $n \ne 1$ gauge
theories.

\section{Acknowledgment} The work of DS was support by a 2003-2004
Fulbright Fellowship. DS would like to thank N. Ohta for bringing 
several related works to his attention, and M. Rausch de Traubenberg 
for valuable discussions.

\end{document}